\documentstyle[12pt,epsfig,cite]{article}
\newcommand{\be}{\begin{equation}}
\newcommand{\ee}{\end{equation}}
\newcommand{\bea}{\begin{eqnarray}}
\newcommand{\eea}{\end{eqnarray}}

\newcommand{\la}{\langle}
\newcommand{\ra}{\rangle}

\newcommand{\p}{\partial}

\newcommand{\SL}[0]{SL(2,R)}

\def\CF{{\mathcal{F}}}
\def\CP{{\mathcal{P}}}
\def\CQ{{\mathcal{Q}}}
\def\IB{\relax\hbox{$\inbar\kern-.3em{\rm B}$}}
\def\IC{\relax\hbox{$\inbar\kern-.3em{\rm C}$}}
\def\ID{\relax\hbox{$\inbar\kern-.3em{\rm D}$}}
\def\IE{\relax\hbox{$\inbar\kern-.3em{\rm E}$}}
\def\IF{\relax\hbox{$\inbar\kern-.3em{\rm F}$}}
\def\IG{\relax\hbox{$\inbar\kern-.3em{\rm G}$}}
\def\IGa{\relax\hbox{${\rm I}\kern-.18em\Gamma$}}
\def\IH{\relax{\rm I\kern-.18em H}}
\def\IK{\relax{\rm I\kern-.18em K}}
\def\IL{\relax{\rm I\kern-.18em L}}
\def\IP{\relax{\rm I\kern-.18em P}}
\def\IR{\relax{\rm I\kern-.18em R}}
\def\IZ{\relax{\rm Z\kern-.5em Z}}

\def\half{\frac{1}{2}}
\def\f{\frac}
\def\p{\partial}
\def\k{\kappa}

\begin{document}

\begin{titlepage}

\rightline{February 2001}

\vskip 2 cm

\begin{center}
{\LARGE Logarithmic Currents in the $SU(2)_0$ WZNW model}
\vskip 1 cm

{\large A. Nichols\footnote{a.nichols1@physics.ox.ac.uk}}

\vskip 1.5 cm

{\it Theoretical Physics, Department of Physics, University of Oxford}\\
{\it 1 Keble Road, Oxford, OX1 3NP,  UK}

\end{center}

\begin{abstract}

We study four point correlation functions of the spin $1$ operators in the $SU(2)_0$ WZNW model. The general solution which is everywhere single-valued has logarithmic terms and thus has a natural interpretation in terms of logarithmic conformal field theory. These are not invariant under all the crossing symmetries but can remain if fields possess additional quantum numbers.

\end{abstract}

\end{titlepage}

\newpage

\section{Introduction}

Conformal field theory (CFT) \cite{Belavin:1984vu} has had immense success both in describing critical phenomena and also in its applications to string theory. Logarithmic terms in correlation functions were first observed in \cite{Rozansky:1992rx}. It was realized that these were due to the indecomposable representations that can occur in the OPE of primary fields \cite{Gurarie:1993xq}. Since then there has been a large amount of literature in discovering new examples and applications \cite{Bilal:1994nx,Caux:1996nm,Kogan:1996wk,Maassarani:1996jn,Kogan:1996wk,CTT,Kogan:1996zv,Ellis:1996fi,RahimiTabar:1997ub,Kogan:1997fd,Kogan:1997vx,Caux:1997kq,Caux:1998eu,Mavromatos:1998nz,Kim:1998yw,Flohr:1998ew,Lewis:1999qv,Kogan:1999hz,Ellis:1999mj,Bhaseen:1999nm,Kogan:1999bn,Kogan:1999hz,Leontaris:2000iq,Nichols:2000mk,Lewis:2000tn,Kogan:2000nw} as well as the general classification \cite{Rohsiepe:1996qj}. It has been shown \cite{Kausch:1995py,Flohr:1997wm,Flohr:2000mc} that correlators obey all the normal properties of a conformal field theory - in particular crossing symmetry and single-valuedness. The process of constructing the full modular invariant conformal field theory from its chiral constituents is more involved than in normal CFT.

The WZNW model is of great importance in CFT. Correlation functions in such models obey differential, Knizhnik-Zamolodchikov (KZ), equations \cite{Knizhnik:1984nr} coming from null states in the theory. The correlation functions of fundamental fields in these models have been calculated explicitly. The solutions to the KZ equations and correlation functions in the $SU(2)_k$ model were studied by \cite{Zamolodchikov:1986bd,Christe:1987cy}. There is a simple Dotsenko-Fateev integral representation for solutions but these do not converge in many cases. In particular in the cases in which logarithms appear we have to be very careful when analytically continuing the solutions.  The factorisation of the KZ equation and the decoupling of null vectors in the unitary $SU(2)_k$ models was studied in general in \cite{Christe:1987yf}.

The $SU(2)_0$ WZNW model was studied in \cite{Kogan:1996df} in the context of describing NS-5 branes. Later $SU(2)_0$, and its supersymmetric extension, were studied in more detail \cite{Caux:1997kq}. It was found that there were logarithmic terms in the four point correlation function of the fundamental spin $\half$ primaries. Here we extend this to the case of spin one operators. We shall see that although there are single-valued logarithmic solutions they do not satisfy all the additional constraints coming from crossing symmetry. The $SU(2)_0$ model is also interesting for studies of $OSp(4|4)_1$ \cite{BernardLC,Bhaseen:2000bm,Bhaseen:2000mi}.

Recently it has suggested that the stress tensor itself might have a logarithmic partner. This was originally proposed in a $c=0$ CFT \cite{Gurarie:1999yx}, which can be non-trivial if it is not required to be unitary. It was later realised that the same situation could occur for general central charge \cite{Moghimi-Araghi:2000qn, Flohr:2000mc}. This is due to the fact that the identity has a logarithmic partner and thus it becomes a zero norm state \cite{Caux:1996nm}. In the WZNW models the stress tensor is given by the Sugawara construction and it would be interesting to see if the Kac-Moody currents also had logarithmic partners and if the new spin 2 fields could be constructed from these \cite{Susy30,WorkInProgress}.
\section{Correlation functions}

The WZNW theory has affine $SU(2)$ left and right moving symmetries whose modes generate the Kac-Moody algebra:
\be
~[ J^a_n,J^b_m ]  =  i {f^{ab}}_c J^c_{n+m} + k n \delta^{ab} \delta_{n+m,0} 
\ee

We have similar expressions for the $\bar{J}^a_n$. We introduce the following representation for the  $\SL$ generators \cite{Zamolodchikov:1986bd}:
\bea \label{eq:repn}
J^+=x^2\frac{\p}{\p x}-2jx, ~~~ 
J^-=-\frac{\p}{\p x}, ~~~
J^3=x\frac{\p}{\p x}-j \eea

The stress-energy tensor in the case of the ungauged WZNW model is given in terms of the normal Sugawara construction:
\be \label{eq:Sugawara}
T(z)=\frac{1}{k+2} :J^a(z) J^a(z):=\frac{1}{k+2} :\half J^+J^- + \half J^-J^+ +J^3 J^3 :
\ee
where : : denotes normal ordering. 
The modes $L_n$ obey the standard Virasoro algebra with central charge $c=\frac{3k}{k+2}$.
We introduce primary fields, $\phi_j(x,z)$  of the KM algebra:
\be
J^a(z)\phi_j(x,w) = \frac{1}{z-w} J^a(x) \phi_j(x,w) 
\ee
where $J^a(x)$ is given by (\ref{eq:repn}). The fields $\phi_j(x,z)$ are also primary with respect to the Virasoro algebra with $L_0$ eigenvalue:
\be
h=\frac{j(j+1)}{k+2}
\ee

Two and three point functions are determined up to a multiplicative constant.
For the case of the four point correlation functions of $SU(2)$ primaries the form is determined by global conformal and $SU(2)$ transformations up to a function of the cross ratios.
\bea \label{eq:correl} 
\langle \phi_{j_1}(x_1,z_1) \phi_{j_2}(x_2,z_2) \phi_{j_3}(x_3,z_3) \phi_{j_4}(x_4,z_4) \rangle
&=&z_{43}^{h_2+h_1-h_4-h_3}z_{42}^{-2h_2}z_{41}^{h_3+h_2-h_4-h_1} \nonumber \\
& & z_{31}^{h_4-h_1-h_2-h_3}x_{43}^{j_3+j_4-j_1-j_2}x_{42}^{2j_2} \nonumber \\
& & x_{41}^{j_1+j_4-j_2-j_3}x_{31}^{j_1+j_2+j_3-j_4}~F(x,z)
\eea 
Here the invariant cross ratios are:
\be 
x=\frac{x_{21}x_{43}}{x_{31}x_{42}} ~~~ z=\frac{z_{21}z_{43}}{z_{31}z_{42}} 
\ee

In a normal CFT we expect the OPE of two primary fields to take the general form:
\be
\label{eq:OPE}
\phi_{j_1}(x_1,z_1) \phi_{j_2}(x_2,z_2) = \sum_J{C(j_1,j_2,J) z_{12}^{-h_1-h_2+h_J} x_{12}^{j_1+j_2-J} [ \phi_J(x_2,z_2) ] } 
\ee
where we have denoted by $[ \phi_J ]$ all descendent fields that can be produced from the given primary field $ \phi_J $. In principle given $C(j_1,j_2,j_3)$, we know the entire operator content of the theory and should be able to determine all higher point correlation functions using the OPE (\ref{eq:OPE}) and the crossing symmetries. This is called conformal bootstrap and has only so far been solved for the minimal models. We will see however that our solutions require more operators to be included in the OPE. 

Correlation functions of the WZNW model satisfy a set equations known as the Knizhnik-Zamolodchikov (KZ) equations due to constraints from the null states following from (\ref{eq:Sugawara}). These are:
\be
|\chi \ra = ( L_{-1} - \frac{2}{k+2} J^a_{-1}J^a_0 ) |\phi \ra
\ee

For two and three point functions this gives us no new information. However for the four point function (\ref{eq:correl}) it becomes a partial differential equation for $F(x,z)$. In contrast to the case of \cite{Kogan:1999hz,Nichols:2000mk,Lewis:2000tn} we are dealing purely with the finite dimensional representations of $SU(2)$.
If we now use our representation (\ref{eq:repn}) we find the KZ equation for four point functions is:
\be \label{eq:KZ}
-(k+2) \frac{\p}{\p z} \CF(x,z)=\left[ \frac{\CP}{z}+\frac{\CQ}{z-1} \right] \CF(x,z)
\ee
Explicitly these are:
\bea
\CP \!\!\!\!&=&\!\!-x^2(1-x)\frac{\p^2}{\p x^2}+((-j_1-j_2-j_3+j_4+1)x^2+2j_1x+2j_2x(1-x))\frac{\p}{\p x} \nonumber \\
& & -2j_2(-j_1-j_2-j_3+j_4)x-2j_1j_2 \\
\CQ \!\!\!\!&=&\!\!-(1-x)^2x\frac{\p^2}{\p x^2}-((-j_1-j_2-j_3+j_4+1)(1-x)^2+2j_3(1-x)+2j_2x(1-x))\frac{\p}{\p x} \nonumber \\
& & +2j_2(j_1+j_2+j_3-j_4)(1-x)-2j_2j_3 
\eea
\section{Doublet solutions}

Before detailing our spin one solutions at $k=0$ we demonstrate that this reproduces the previous ones for the doublet ($j=\half$) representation.
We therefore put $j_i=\half$. Then we get:
\bea
\CP &=& x^2(x-1)\frac{\p^2}{\p x^2}+(2x-x^2)\frac{\p}{\p x}+(x-\half) \\
\CQ &=& -(1-x)^2x\frac{\p^2}{\p x^2}+(x^2-1)\frac{\p}{\p x}+(\half-x) 
\eea

To reduce this partial differential equation to a set of ordinary ones for $z$ we impose the fact that the representations are really the irreducible $j=\half$ ones. This leads to the ansatz:
\be
\CF(x,z)=A(z)+xB(z)
\ee

Substituting this into the KZ equation and separating powers of $x$ leads to:
\bea
\frac{\p A}{\p z}&=&\frac{-1}{(k+2)}\frac{1}{z(z-1)}\left[-zB(z)+\half A(z)\right] \\
\frac{\p B}{\p z}&=&\frac{-1}{(k+2)}\frac{1}{z(z-1)}\left[-A(z)+2zB(z)-\frac{3}{2}B(z)\right]
\eea

Now letting $F_1=A+B \quad F_2=-A$ we obtain the standard form \cite{YellowPages}.
%
%
\bea
\frac{\p F_1}{\p z}&=&\frac{-1}{k+2}\left[\frac{3 F_1}{2z}+\frac{F_2}{z}-\frac{F_1}{2(z-1)}\right] \\
\frac{\p F_2}{\p z}&=&\frac{-1}{k+2}\left[\frac{3 F_2}{2(z-1)}+\frac{F_1}{z-1}-\frac{F_2}{2z}\right]
\eea

For generic $k$ these have linearly independent hypergeometric solutions \cite{Knizhnik:1984nr}. However for certain values of $k$ the solutions develop logarithmic singularities.
For $k=0$ these have the solutions \cite{Caux:1997kq}:
\bea
F^a_1&=&-\f{1}{2} F\left(\f{1}{2},\f{3}{2};1;z\right) ~~~
F^a_2=-\f{1}{4} F\left(\f{1}{2},\f{3}{2};2;z\right) \\
F^b_1&=&\f{\pi}{4} F\left(\f{1}{2},\f{3}{2};2;1-z \right) ~~~
F^b_2=-\f{\pi}{2} F\left(\f{1}{2},\f{3}{2};1;1-z \right)
\eea

The full correlator can be constructed from these by imposing single-valuedness and crossing symmetry and one finds that the logarithmic terms remain.
For this case there is little value in using the auxiliary variable approach but for higher spin representations it allows easy calculation of the matrices $\CP$ and $\CQ$.
\section{Spin 1 Triplet representation}

For $j=1$ we get:
\bea
\CP &=& x^2(x-1)\frac{\p^2}{\p x^2}+(-3x^2+4x)\frac{\p}{\p x}+(4x-2) \\
\CQ &=& -(1-x)^2x\frac{\p^2}{\p x^2}+(3x^2-2x-1)\frac{\p}{\p x}+(2-4x) 
\eea
We now impose:
\be
\CF(x,z)=F(z)+xG(z)+x^2H(z)
\ee
Proceeding as before we now get three equations:
\bea \label{eq:3eqns}
\k\f{d F}{d z}&=&\f{-zG(z)+2F(z)}{z(z-1)} \nonumber \\
\k\f{d G}{d z}&=&\f{-4F(z)+2zG(z)-2G(z)-4H(z)z}{z(z-1)} \\
\k\f{d H}{d z}&=&\f{6H(z)z-4H(z)-G(z)}{z(z-1)} \nonumber
\eea
where $\k=-k-2$.

These lead to third order ODEs for $F,G$ and $H$. We calculate only the equation for $F$ as $G,H$ can be obtained easily from $F$.
\bea \label{eq:Feqn}
& &-\k^3 z^3 (z-1)^3 \f{d^3 F(z)}{d z^3}+ \k^2 z^2 (z-1)^2 \left( (8-3\k)z-4 \right) \f{d^2 F(z)}{d z^2} \\
& &+ \k z(z-1) \left( (\k-2)(6-\k)z^2-6(\k-2)z+(2\k+4) \right) \f{d F(z)}{d z} \nonumber \\
& &+ \left( 2 z^2\k(\k+6)-2z(3\k+8)(\k+2)+4(\k+2)^2 \right) F(z) \nonumber
\eea

In \cite{Fuchs:1987ew} the four point function of vector fields in the  $O(N)$ model was given by an integral representation for general $k$. Our calculation is similar and has the integral representation:
\be
F(z)= \int_{C_1} {ds \int_{C_2} {dt ~ (s t)^{\alpha} ((s-1)(t-1))^{\beta} ((s-z)(t-z))^{\gamma} (s-t)^{\delta} }}
\ee
where $\alpha=\f{2}{\k} ~~~~ \beta=\f{2}{\k}-1 ~~~~ \gamma=\f{2}{\k} ~~~~ \delta=-\f{2}{\k}$.

However for for certain values of $\k$, or choices of contours, this fails to converge and is not a very useful form. 
We did not manage to solve (\ref{eq:Feqn}) completely but did get some simple solutions at $k=0$ ($\k=-2)$.
This is based on the factorization of the above equation that occurs. We find we can rewrite the above equation as:
\bea
\left( (z-1)\f{d}{d z}+2 \right)
\left( z(z-1)\f{d}{d z}-1 \right)
\left( (z-1)\f{d}{d z}+1 \right) F(z)=0
\eea

We can easily solve this and hence find solutions for $F(z)$ and thus obtain $G,H$ from them. Combining into a solution for $\CF(x,z)$ these can be conveniently written in a basis:
\bea
\CF_1(x,z)&=&-\f{1}{2(z-1)}+\f{x}{z}+\f{x^2}{2z(z-1)} \\
\CF_2(x,z)&=&-\f{1+(z-1)(\ln(1-z)-\ln(z))}{2(z-1)^2}+\f{x(z+(z-1)^2(\ln(1-z)-\ln(z)))}{z(z-1)^2} \nonumber \\ 
    & & + \f{x^2(1-2z+z(z-1)(\ln(1-z)-\ln(z)))}{2z^2(z-1)^2}  \\
\CF_3(x,z)&=&-\f{1-z+\ln(z)}{2(z-1)}+\f{x\ln(z)}{z}+\f{x^2(1-z+z\ln(z))}{2z^2(z-1)}
\eea

It is easily verified that these solve the above equations (\ref{eq:3eqns}). As we are dealing with the finite dimensional representations we wish to remove the $x$ dependence at the end by expanding in terms of $x_1,x_2,x_3,x_4$. However it is generally more convenient to do so after imposing the constraints due to crossing symmetry.
We now combine these with their antiholomorphic components into the full correlator
\be
G(x,\bar{x},z,\bar{z})=\sum_{a,b=1}^{3}{U_{a,b} \CF_a(x,z) \overline{\CF_b(x,z)}}
\ee

To make this single-valued everywhere we find:
\bea \label{eq:Gfull}
G(x,\bar{x},z,\bar{z})&=&U_{1,1} \CF_1(x,z) \overline{\CF_1(x,z)} 
+ U_{1,2} \Bigl[ \CF_1(x,z) \overline{\CF_2(x,z)} + \CF_2(x,z) \overline{\CF_1(x,z)} \Bigr] \nonumber \\
& &+ U_{1,3} \Bigl[ \CF_1(x,z) \overline{\CF_3(x,z)} + \CF_3(x,z) \overline{\CF_1(x,z)} \Bigr]
\eea

In contrast to normal CFT this does not have a simple diagonal form. This is one well known difference in logarithmic CFT. In order to get a well defined correlator we must also impose the crossing symmetries:
\bea
& &\label{eq:cross1} G(x,\bar{x},z,\bar{z})=G(1-x,1-\bar{x},1-z,1-\bar{z}) \\
& &\label{eq:cross2} G(x,\bar{x},z,\bar{z})=z^{-2h}\bar{z}^{-2h}x^{2j}\bar{x}^{2j} G(\f{1}{x},\f{1}{\bar{x}},\f{1}{z},\f{1}{\bar{z}}) 
\eea
Under $x,z \rightarrow 1-x,1-z$ we have :
\be
\CF_1 \rightarrow \CF_1 \quad \CF_2 \rightarrow -\CF_2 \quad \CF_3 \rightarrow \CF_2 + \CF_3
\ee
Under $x,z \rightarrow \f{1}{x},\f{1}{z}$:
\be
\CF_1 \rightarrow \f{z^2}{x^2}\CF_1 \quad \CF_2 \rightarrow \f{z^2}{x^2}(i\pi \CF_1+\CF_2+\CF_3) \quad \CF_3 \rightarrow -\f{z^2}{x^2}\CF_3
\ee

If we put $U_{1,3}=2~U_{1,2}$ then our solution is invariant under (\ref{eq:cross1}) whereas if $U_{1,2}=2~U_{1,3}$ then we have invariance under (\ref{eq:cross2}). Thus the only solution obeying both crossing symmetries is $U_{1,2}=U_{1,3}=0$ and so the logarithmic solutions do not contribute to the correlator. If there is some other symmetry present in a problem, besides the affine $SU(2)_0$ then the correlator will not be invariant under all the crossing symmetries. We comment on this later.  
We can now remove the $x$ dependence and go back to the {\it{index}} notation. We do this using
\bea
\phi_{1}(x,z)=x^2 \phi^{+}_{1}(z) + x \phi^{0}_{1}(z) +  \phi_{1}^{-}(z)
\eea
with similar expressions for the antiholomorphic part. From the general form in auxiliary variables we get the following expected two and three point equations:
\be 
\la \phi^{a}_{1}(x_1,z_1) \phi^{b}_{1}(x_2,z_2) \ra = A \frac{\eta_{ab}}{z_{12}^{2}}
\ee
\be
\la \phi^{a}_{1}(x_1,z_1) \phi^{b}_{1}(x_2,z_2) \phi^{c}_{1}(x_3,z_3) \ra = \frac{B~\epsilon^{abc}}{z_{12} z_{13} z_{23}} \nonumber
\ee
We can now do the same for our four point function. If $U_{1,2}=U_{1,3}=0$ in (\ref{eq:Gfull}) then we have the correlator:
\bea  \label{eq:allsame}
&&\langle \phi^{a \bar{a}}_{1}(z_1) \phi^{b \bar{b}}_{1}(z_2) \phi^{c \bar{c}}_{1}(z_3) \phi^{d \bar{d}}_{1}(z_4) \rangle =\\
&& C \Bigl[ \frac {\delta^{ab}\delta^{cd}}{z_{13}z_{14}z_{23}z_{24}} + \frac {\delta^{ac}\delta^{bd}}{z_{12}z_{34}z_{14}z_{32}} + \frac {\delta^{ad}\delta^{bc}}{z_{13}z_{12}z_{43}z_{42}} \Bigr] \overline{\Bigl[ \frac {\delta^{ab}\delta^{cd}}{z_{13}z_{14}z_{23}z_{24}} + \frac {\delta^{ac}\delta^{bd}}{z_{12}z_{34}z_{14}z_{32}} + \frac {\delta^{ad}\delta^{bc}}{z_{13}z_{12}z_{43}z_{42}} \Bigr] } \nonumber
\eea
$A,B$ and $C$ are constants which can only be obtained through the consistency of all the OPE's. In unitary CFT we can always normalize $A=1$ as all null operators must decouple. In our case $k=0$ and even the Kac-Moody current $J(z)$ and the stress tensor $T(z)$ have vanishing two point functions, and consequently we cannot impose this if we wish to obtain a non-trivial theory.
Assuming that $C \ne 0$ then expanding the four point function above reveals:
\bea \label{eq:j11OPE}
&&\phi^a(z) \phi^b (0) \sim \f{i \epsilon^{abc}Q^c}{z} + \delta^{ab}R(0) + ... \\
&&Q^a(z) Q^{b}(0) \sim \f{1}{z^2} \\
&&R(z) R(0) \sim \f{1}{z^4}
\eea

We have also found explicit solutions for the case of $j_1=j_3=1 ~~~ j_2=j_4=\half$. In this case we have only two conformal blocks. We obtain two solutions, one of which has logarithmic terms in its expansion, the other being well behaved:
\bea
\CF (x,z) = && A \Bigl[ \f{\arcsin(2z-1)+2 \sqrt{z(1-z)}}{z-1} + x \f{(2z-1)\arcsin(2z-1)+2\sqrt{z(1-z)}}{z(1-z)} \Bigr] \nonumber \\
&&+ B \Bigl[ - \f{1}{1-z} + x \f {2z-1}{z(1-z)} \Bigr] 
\eea

Imposing the crossing symmetry $x,z \rightarrow 1-x,1-z$ implies $A=0$ and we are again left with the non-logarithmic terms. As previously mentioned the OPEs for two fundamental fields were found from the four point function \cite{Caux:1997kq}. By expanding the above expression one can combine our results (\ref{eq:j11OPE}) with theirs.

If we have additional global symmetries present \cite{Kogan:1999hz} and fields have extra quantum numbers then it is not necessary to impose all of the crossing symmetry. Then we find that we may have logarithmic solutions:
\bea \label{eq:bosonic}
\langle B_1(x_1,z_1) B_2(x_2,z_2)B_1(x_3,z_3) B_2(x_4,z_4) \rangle =U_{1,2}\Bigl[ F_1(x,z) \overline{ (F_2(x,z)+2F_3(x,z)) } \\
+  (F_2(x,z)+2F_3(x,z)) \overline{F_1(x,z)} \Bigr] \nonumber
\eea
\bea
\langle F_1^{a \bar{a}}(z_1,\bar{z_1}) F_2^{b \bar{b}}(z_2,\bar{z_2}) F_1^{c \bar{c}}(z_3,\bar{z_3}) F_2^{d \bar{d}}(z_4,\bar{z_4}) \rangle =\Biggl\{ \delta^{ab}\delta^{cd}\Big[\f{1}{z_{12}^2z_{34}^2}+\f{\ln(\f{z_{12}z_{34}}{z_{23}z_{14}})}{z_{23}z_{14}z_{13}z_{24}} \Bigl] \\
+ \delta^{ac}\delta^{bd}\Big[\f{\ln (\f{z_{23}z_{14}}{z_{12}z_{34}} )}{z_{12}z_{34}z_{23}z_{14}} \Bigl] - \delta^{ad}\delta^{bc}\Big[\f{1}{z_{23}^2z_{14}^2}+\f{\ln(\f{z_{23}z_{14}}{z_{12}z_{34}} )}{z_{12}z_{34}z_{13}z_{24}} \Bigl] \Biggl\} \times c.c \nonumber 
\eea
where we have denoted bosonic fields by $B_{1,2}$ and fermionic ones by $F_{1,2}$. In (\ref{eq:bosonic}) we have ignored the contribution of terms of the type (\ref{eq:allsame}) as these have already been commented on.

In the fermionic case we have the following OPE's.
\bea
F^a_1(z_1)F^b_2(z_2) &\sim& \f{\delta^{a b}}{z_{12}^2} + \f{i \epsilon^{abc}P^c (z_2)}{z_{12}} + \f{\ln z_{12}~ i \epsilon^{abc}  Q^c(z_2)}{z_{12}} \\
&& + i \epsilon^{abc}R^e(z_2) + \ln z_{12}~ i \epsilon^{abc}S^e(z_2) + \delta^{ab}T(z_2)+\ln z_{12} ~ \delta^{ab}t(z_2) + ... \nonumber
\eea

\section{Conclusion}
We have found the general solution for the four point function of the vector representation in the $SU(2)_0$ model. If we impose single-valuedness and all the crossing symmetries then we find that all the logarithmic terms vanish. The logarithmic terms can thus only be present if fields possess additional quantum numbers.

In contrast to irreducible representations in a general indecomposable representation there is no reason to assume that the central extensions must act diagonally. If the logarithmic partner of the stress tensor also involves a non-diagonal central charge then there seems to be no reason that a similar phenomenon could not occur in the Kac-Moody algebra. It would be interesting to see study the relation between these.

In \cite{Zamolodchikov:1986bd} a connection was derived between $SU(2)_k$ WZNW and the minimal models (See also \cite{Bhaseen:1999nm} where it was used to relate the quantum theory of the plateau transition to Liouville). This relates four point functions in $SU(2)_0$ to five point ones in the $c=-2$ minimal model. Presumably this means much of the structure is the same. The $c=-2$ theory is one of the best understood LCFTs \cite{Kausch:1995py} and in particular it is quasi-rational i.e fields fall into a finite number of representations of a higher spin algebra. It would also be interesting to see if the factorization properties of the KZ equation are related to those of the minimal models.
We hope to study some of these questions in future work.

\section{Acknowledgments}
I would like to thank I.I. Kogan, J. Bhaseen, and Y. Ishimoto  for interesting and helpful discussion. This work was funded by PPARC studentship number PPA/S/S/1998/02610. 
%
%

\end{document}